\documentstyle[twocolumn,aps,epsfig]{revtex}
\begin{document}
\draft
\title{Electron Neutrino Sources from the Core of the Earth}

\author{A. Widom and E. Sassaroli}
\address{Physics Department, Northeastern University, 
Boston, Massacusetts, U.S.A}
\author{Y.N. Srivastava }
\address{Physics Department \& INFN, University of Perugia, Perugia, Italy}
\maketitle

\begin{abstract}
The physical interpretation of extensive measurements of electron 
neutrinos (in laboratories located on or somewhat below the Earth's 
surface) often require geophysical notions concerning the possible 
neutrino sources. Here, we discuss the notion that the Earth's core   
is a substantial source of low energy electron neutrinos.   
\end{abstract}  

\narrowtext

Our knowledge about the internal geophysical structure of the Earth has been 
summarized in the classic work of Jeffreys\cite{1}. In his work can be found 
the basic physics of how the structured spherical shells of our planet have  
been deduced. One employs the measured sound wave propagation due 
to the seismic crunching and crackling of Earth quakes. The standard 
geophysical model\cite{2,3} of the Earth is pictured (approximately to scale) 
below in Fig.1. 
\begin{figure}[htbp]
\begin{center}
\mbox{\epsfig{file=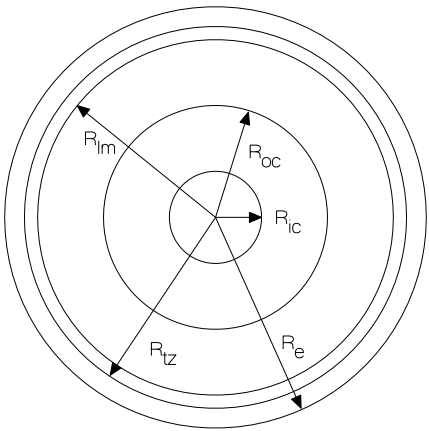,height=70mm}}
\caption{
The shell structure of the Earth includes an inner core with radius 
$R_{ic}$, an outer core  with radius $R_{oc}$, a lower 
mantle with radius $R_{lm}$, a transition zone with radius $R_{tz}$ 
and a thin crust to the surface at radius $R_e$.
}
\label{snufig1}
\end{center}
\end{figure}

Less well developed are our notions of radioactive processes within the 
core and shell structure of the Earth. Early discussions on the nature 
of nuclear physics within the Earth are due to Darwin\cite{4} and 
Rutherford\cite{5}. Later Jeffreys states\cite{6} ``... it would be 
interesting to consider what would happen if the present radioactive 
elements, with their quantities adapted to $4\times 10^9$ years ago, 
were uniformly distributed ..., it is quite possible that radioactive 
heating could produce fusion in a fraction of the age of the Earth.''

Just as the solar neutrino flux \cite{7} constitutes 
{\em direct evidence} of nuclear reactions within the Sun, an 
observed geophysical neutrino flux would constitute {\em direct 
evidence} of nuclear reactions within the Earth. Experimental 
geophysical data will exist in laboratories located on or somewhat below 
the Earth's surface with large fiducial volume neutrino detectors. 
The idea is to measure the differential neutrino flux $d^2\Phi $ (per unit 
time per unit area) within a neutrino energy interval $dE$ and incident 
from a solid angle direction within $d\Omega $. If $\theta $ is the angle 
between a line drawn from the laboratory to the center of the earth and 
the direction of the solid angle $d\Omega $, and if there exist  
spherically symmetric sources, then the differential flux has 
the functional form 
\begin{equation}
\Big({d^2\Phi \over dE d\Omega }\Big)={\cal F}(E,\cos \theta ).
\end{equation} 
If $\eta_e (r,E)d^3{\bf r}$ denotes the number of neutrinos produced within 
an earth volume $d^3{\bf r}$ per unit time with an energy less than $E$, then 
the flux of geophysical neutrinos seen in a laboratory located on  
the Earth's surface is described (in an angular range 
$0< \theta < \pi /2$) by  
$$
{\cal F}_e(E,\cos \theta )=
$$
\begin{equation}
R_e\int_0^{2\cos \theta }
\Big(
{d\eta_e (r=R_e\sqrt{1+x^2-2x\cos\theta},E)\over dE}
\Big)dx. 
\end{equation}

Under the assumption that $\eta_e (r,E)$ is proportional to the mass density 
$\rho_e (r)$ in the Earth, it is possible to plot numerically the angular 
distribution of geophysical neutrinos $(d\Phi_e/d\Omega )$. We have 
carried out such a calculation as shown in Fig.2 below. The mass 
density in the numerical integrals are taken from tables provided by 
Birch\cite{8}. For the purpose of conversion into a neutrino flux, 
we have assumed one part per million of the nuclei $\beta $-decay with a 
mean life-time of $4\times 10^9$ years. These nuclear physics numbers are 
in reasonable agreement with the estimates of Jeffreys. 

We find (using the above estimates) that in an Earth bound laboratory  
the total geophysical electron neutrino flux is at least comparable 
to the total solar electron neutrino flux. The theoretical 
angular distribution of the geophysical neutrino flux is (of course) 
broader. Nevertheless, the magnitudes are sufficient to imply 
some periodic modulations in what has previously been regarded as a 
purely solar neutrino flux. Modulations will occur whenever the neutrino 
beam from the Sun is parallel (or anti parallel) to the neutrino beam 
from the Earth's core.

\begin{figure}[htbp]
\begin{center}
\mbox{\epsfig{file=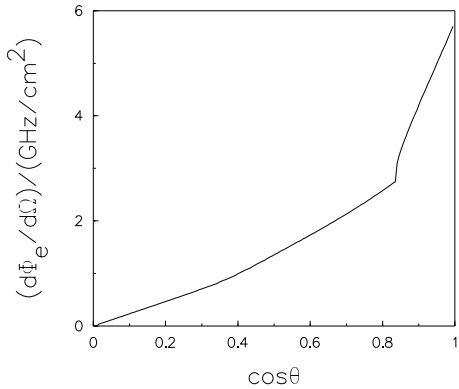,height=70mm}}
\caption{
The angular distribution of geophysical neutrinos in a model in 
which the source intensity of neutrino production is proportional 
to the mass density. The geophysical mass density used is due to Birch[8].
}
\label{snufig2}
\end{center}
\end{figure}

The peak in $(d\Phi_e /d\Omega )$ as $\cos\theta \to 1$ is 
due to the fact that the core of the Earth contains roughly half of 
the Earth's mass and thereby half of the Earth's neutrino sources in 
the model under consideration. The discontinuity of the mass density 
at the core radius $R_{oc}$ is clearly visible from the structure of 
the peak. 

While the nuclear physics estimates within the core of the Earth 
are not as accurate as the nuclear physics estimates within the core 
of the Sun, the above considerations may be valid at least 
in order of magnitude, and are surely worthy of further study. Recent 
geophysical investigations\cite{9} indicate a substantial concentration 
of Uranium (and other radio active nuclei) in the Earth's core, since  
Uranium is totally miscible at the elevated temperatures of the Earth's 
core material. This recent geophysical work supersedes an earlier 
miscibility gap argument for the bulk of nuclear reactions to occur 
in the mantle and/or above the mantle. Further discussions may be 
found in the literature\cite{10}. Given that nuclear reactions take 
place in both the core and mantle, as well as on the surface of the Earth, 
it should be of no surprise that {\em nature} and {\em not nuclear 
physicists} built the first nuclear reactors. These naturally occurring 
nuclear reactors have been relatively recently discovered\cite{11}. 

Atmospheric neutrinos also require geophysical reasoning to trace their 
source. It has been assumed that cosmic ray protons first produce pions 
which then  decay via  
\begin{equation}
\pi^+\to \mu^++\nu_\mu\to e^++\nu_e +\bar{\nu}_\mu+\nu_\mu,
\end{equation}
or
\begin{equation}
\pi^-\to \mu^-+\bar{\nu}_\mu\to e^-+\bar{\nu}_e +\nu_\mu+\bar{\nu}_\mu.
\end{equation}
From Eqs.(2) and (3), and in the absence of neutrino oscillations, one 
expects\cite{12} from atmospheric neutrinos a ratio of 
\begin{equation}
\Big({\Phi_a(\nu_\mu+\bar{\nu}_\mu)
\over \Phi_a(\nu_e+\bar{\nu}_e)}\Big)\approx 2.
\end{equation}
Initial atmospheric neutrino experiments\cite{13,14} were aimed at 
deducing neutrino oscillation magnitudes via deviations from Eq.(5).  

The notion of geophysical neutrino sources within the Earth's core 
was ignored in these experiments. However,  an ``excess'' (above and 
beyond the factor of two) of low energy electron neutrinos heading 
upward from the earth has been observed. In the most recent Super Kamiokande
\cite{15} experiments, the electron neutrino excess was quite pronounced. 
Note, in this regard, that our $\theta $ is related to 
$\theta_{\cal SK}$ used in the Super Kamiokande experiment 
by $\theta =\pi -\theta_{\cal SK}$; i.e. 
\begin{equation}
\cos\theta =-\cos\theta_{\cal SK}.
\end{equation}
If we consider neutrinos with energy $E<0.4GeV$, then we 
estimate (from the excess electron neutrino data at angles 
$0.6<\cos\theta=(- \cos\theta_{\cal SK})<1.0$) the ratio
\begin{equation}
\Big({\Phi_e(\nu_e,\cos\theta > 0.6\ ,E<0.4GeV)\over 
\Phi_a(\nu_e,\cos\theta > 0.6\ , E<0.4GeV)}\Big)\approx 0.3,
\end{equation}
wherein the excess electron neutrinos are here presumed to be of geophysical 
origin from within the Earth's core.

We hope that the hypotheses (and the experimental data) discussed here  
concerning geophysical electron neutrino sources within the Earth's core, 
will inspire future investigations. It appears within current neutrino 
detection technology to map, via Eqs.(1) and (2), geophysical nuclear 
reaction sources inside the Earth's core. Such a technique appears presently 
as a unique probe of such distributions.

\end{document}